\documentclass[a4paper,11pt]{article}
\usepackage{amsmath, amsthm, amsfonts, amssymb, mathrsfs}
\usepackage{subfigure}
\usepackage[dvips]{graphics}
\usepackage{graphicx}

\begin{document}

\title{Electronic transport through the phtalocyanine molecule in atomic contacts Co}           

\author{$^{a}$Ali Jaafar,
\footnote{E-mail address: alijaafar.ja@gmail.com}
$^{b}$Tarek Khalil
\footnote{E-mail address: tkhalil@ul.edu.lb}}
\maketitle
\begin{center}
$^{a}$  Laboratoire de Physique des Mat\'eriaux, Universit\'e Libanaise, Facult\'e des Sciences (I), Hadath, Beyrouth, Liban\\
$^{b}$ Department of Physics, Faculty of Sciences(V), Lebanese  University, Nabatieh, Lebanon\\
\end{center} 
 

\begin{abstract} 
The effect of magnetic STM-tip on electronic, magnetic and electronic transport properties
through the molecule junction  STM-tip-Co/CoPc/Co(111), has been investigated by mean of
$ab$ $initio$ electronic structure calculations. The spin transition has been studied by varying 
the distance (passing from the tunneling regime to the contact regime) between the tip and
the CoPc molecule in both configurations, parallel and anti-parallel. Our calculation shows
that the transition of spin of the Co atom of CoPc molecule has led to a change of the
sign of the \textit{Magneto-Resistance} ($MR$). It is also shown that the characteristic \textit{I-V}
has been influenced by this spin-transition of central atom of CoPc molecule.  
\end{abstract}

\maketitle 

\section{Introduction} 
The idea of molecular electronics and spintronics, is one of the most promising
avenues for further miniaturization of devices required. 
In fact, the organic devices must be contacted by inorganic drivers to enable
communication and control from outside. The crucial contact organic-inorganic
can change dramatically the properties of the molecule and determine the functionality
of the device. So, the study of contact between molecules and inorganic, semiconductors
or metallic substrates, is one of the most important tasks in the emerging technology
of molecular electronics \cite{Reed}. In molecular spintronics, the fundamental properties
is especially the magnetism of the electrodes of the molecular junction. In addition for
metal-organic molecules, the magnetism is not limited to the electrodes as the metallic part
can cause different interesting effects such as the Kondo effect. Traditionally, metal
phthalocyanine (XPc) molecules, composed of a metal atom surrounded by a ligand ring,
have been used in this contexte, also because of their potential interest for spin-dependent
electronics \cite{Scheybal,Wende} and optoelectronics \cite{Liu,Elemans}.   
Many researches have been performed on Pc molecules \cite{Zhao,GC,JM,CI,AF,Lan,Arrigo,Xi},
such as the controllable Kondo effect\cite{L.Gao,Zhao,GC,JM} and vibrational spin of individual
CoPc molecules\cite{CI}. The transport through CoPc molecule junction has also been 
discussed\cite{CI,AF,Lan,Arrigo,XiChen}. In general, it has been shown that the electronic 
and magnetic properties of a XPc molecule strongly depend on the type of metal ion within the
phthalocyanine ligand and the type of surface on which the molecule is adsorbed \cite{AF,Javaid}.
Recently, several theoretical and experimental \cite{CI,XiChen} works have been done to study the
influence of magnetic and non-magnetic substrates on the molecule CoPc, as well as the spin
transport across single cobalt-phthalocyanine molecules.
On the other hand, there is now a fast growing interest in controlling the
spin orientation of a single or few magnetic atoms in a solid state environment
for future spintronics and quantum information devices \cite{C.LeGall,M.Goryca}. 
The spin direction can be driven into either parallel or antiparallel
(ferromagnetic or antiferromagnetic) alignment, or even noncollinear alignment,
depending on the sign of the exchange coupling \cite{Bergmann}. Recently, scanning
tunneling microscope (STM) offers the ability to study single magnetic moments and
the exchange coupling between spins in a precisely characterized local environment
\cite{Hirjibehedin1,Wahl}. Besides, the use of STM is able to control the
spin state of a single magnetic adatom, such as by depositing it on
an insulating thin film \cite{Hirjibehedin2} or on magnetic islands \cite{Heinrich,Yayon}. 
But the influence of the STM tip on electronic and magnetic properties of adatoms
has not been investigated in all the above experimental studies. However, it
was clearly investigated in some other theoretical and experimental studies \cite{Tao,Hofer,Neel1,R.Huang,Neel2,Vitali}  
which have shown that the tip-surface distance strongly influences the electronic
and magnetic properties as well as the behavior of the conductance of a single
adatom on metallic surfaces.
Our aim in this theoretical study is to combine the two ideas :

\begin{itemize}
\item[1-] The influence of the magnetic substrate on the transport properties 
through molecule junction, 
\item[2-] The strong impact of the STM-tip on the electronic and magnetic properties
of the molecule junction.
\end{itemize}

Performing {\sl ab initio} pseudopotential calculations we
have studied the magnetic moment of the central atom of CoPc, the transmission coefficients through
a molecular junction, and as well as the I-V characteristics as function of the distance 
between the STM tip and the CoPc molecule deposited on Co(111) surface. 
In fact, the junction adopted in our work consists of two Co(111) electrodes,
a cone of cobalt which represents the magnetic tip of STM and CoPc molecule
deposited with different types of stacking (hcp, cfc, bridge) on the second electrode 
(Co(111)). The Co(111) substrate is modeled by a slab of 5 layers of Co atoms,
and the STM tip consists of 11 atoms.

In order to manipulate the spin direction of the central atom of CoPc and the conductance across the
molecular junction by a magnetic STM tip, we have adopted two configurations (see Fig.1). 
In the first configuration (Fig.1-(a)), the initial spin directions of the STM tip, 
central atom of CoPc and Co(111) substrate are in parallel alignment (ferromagnetic).
Whereas in the second configuration (Fig.1-(b)), the spin directions of tip and CoPc/Co(111) are in antiparallel alignment (antiferromagnetic).

\begin{figure}[htpb]
\centering
\includegraphics[width=11cm, height=8cm]{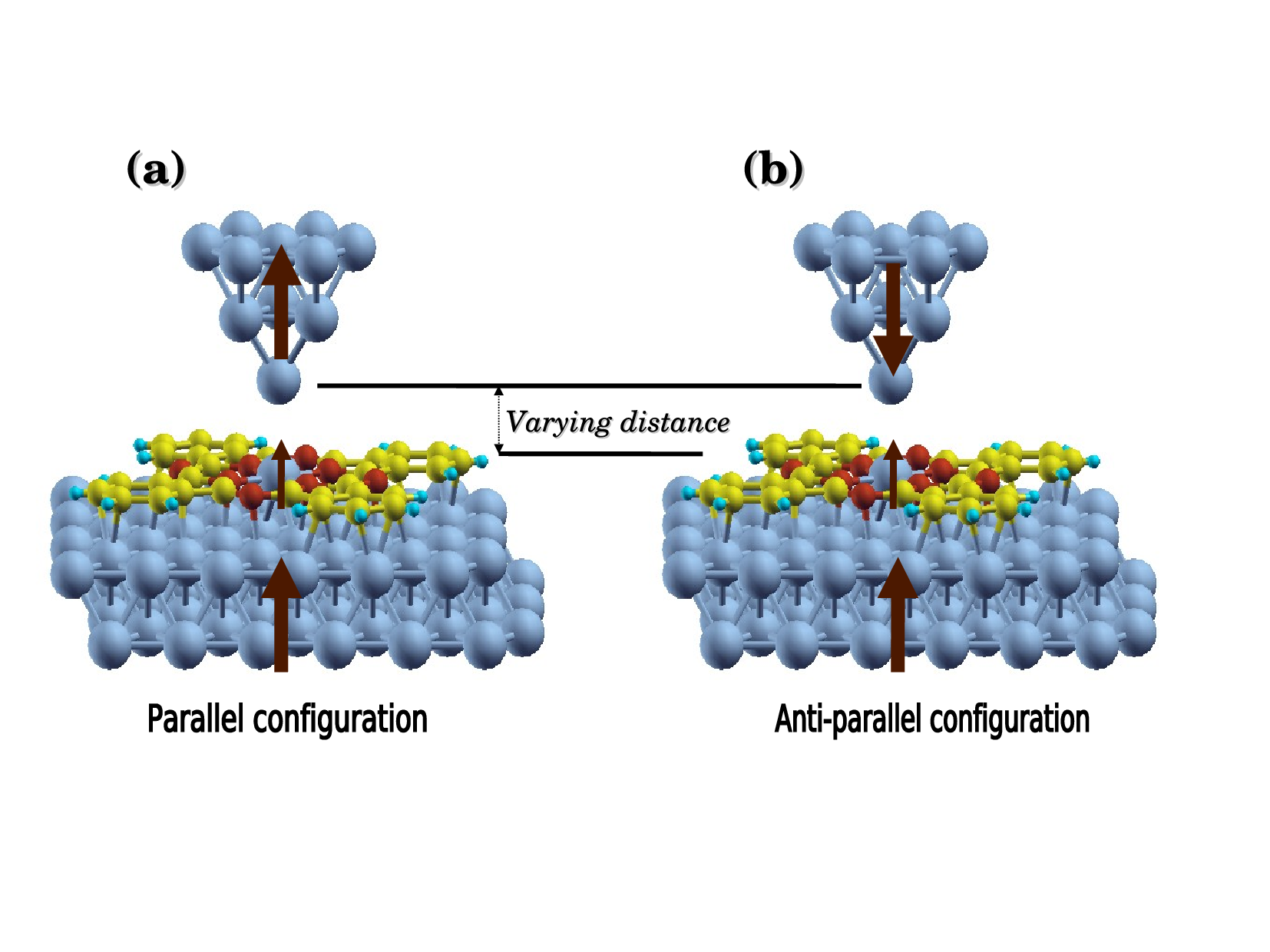}
\label{fig1}
\caption{(a) and (b) represent ferromagnetic and antiferromagnetic configurations respectively. These two figures
show also the varying distance between the STM Co-tip and CoPc molecule deposited on Co(111).}
\end{figure}

Our manuscript is organized as follows: In the second section we describe briefly
our method of calculation. In the third section we present and discuss our results
on the adsorption energy, the electronic and magnetic properties of CoPc/Co(111). 
The effect of the magnetic STM-tip on 
the electronic as well as magnetic properties and the spin-dependent electric 
conductance for the two configurations (ferromagnetic and antiferromagnetic).

\section{Methods of calculation : SIESTA and Smeagol methods }
Our study is based on  {\sl ab initio} pseudopotential calculations using the
{\sl SIESTA} \cite{siesta} and SMEAGOL \cite{smeagol1,smeagol2-Alex} codes.
In fact, our study was performed in two stages. In the first step, we used SIESTA code to determine
the link between structural and electronic properties. In the second step, we use SMEAGOL code
to investigate the electronic transport properties for the two ferromagnetic and antiferromagnetic. \\
SIESTA \cite{siesta} is a calculation method based on the standard Kohn-Sham selfconsistent
DFT which can be used either in the local-density approximation (LDA) \cite{payne} or in the
generalized gradient (GGA) approximation for the electron-electron exchange and correlation
interactions\cite{perdew}. The one-particle problem is solved using linear combination of atomic
orbitals (LCAO) and taking as solution method the diagonalization of the Hamiltonian.
In the presented calculations we have used for the exchange and correlation potential the 
local-density approximation (LDA) of Ceperley-Alder \cite{ceperdley}. For the magnetic systems, 
we performed spin-polarized calculations in the local spin-density approximation (LSDA).
For the ion-electron interactions, the core electrons are replaced by norm-conserving pseudopotentials
\cite{troullier} factorized in the Klienman-Bylander form \cite{kleinman}, including scalar-relativistic
effects. Valence states are described using double-$\zeta$ (DZ) basis sets. For cobalt, a DZ basis was
used with two different radial functions to represent the 4s and 3d orbitals. For N, C and H, a DZ basis 
was used with two different radial functions to represent the 2s, 2p and 1s (for H) orbitals.
Finally for our solid systems, integrations over the Brillouin zone are replaced by converged sums 
over selected $\overrightarrow{k}$ sets \cite{monkhorst}. The electronic population is sorted for each
orbital on each atomic site using the Mulliken analysis implemented in $SIESTA$ method. \\

The $Smeagol$ (Spin and Molecular Electronics on Atomically Generated Orbital Landscape) 
is a fully spin-polarized code \cite{smeagol1,smeagol2-Alex} and an ab initio electronic transport code based on a 
combination of Density Functional Theory (DFT) and Non-Equilibrium Green's function transport methods (NEGF).  
$Smeagol$ uses $SIESTA$ as DFT platform because the final product of $SIESTA$ is a tight-binding like Hamiltonian 
and can be easily interfaced to the NEGF method. The NEGF method splits up a two-terminal device into three regions, 
a scattering and a semi-infinite left/right leads. The bias is applied by setting the chemical potential of the left and 
right leads to $\mu_{L}=\mu+eV_{b}/2$ and $\mu_{R}=\mu-eV_{b}/2$, respectively, where $\mu$ is the common Fermi level
of both leads and $V_{b}$ is the applied bias. The current through the atomic scale system can be calculated from
corresponding Green's function and self-energies \cite{smeagol2-ivan} as follows:   
 
\begin{equation}\label{eqn:current}
I(V) = \frac{2e}{h} \int_{-\infty}^{+\infty} T(E,V)[f_{l}(E-\mu_{l}) - f_{r}(E-\mu_{r})]dE,
\end{equation}
\noindent
where $f_{l}$ and $f_{r}$ are the corresponding electron distribution of the left and right electrodes respectively.
$T(E,V)$ is the transmission coefficient at energy $E$ and bias voltage $V$, which is given by

\begin{equation}\label{eqn:trans}
T(E,V) = Tr[Im \Sigma_{l}(E) G^{R}(E) Im \Sigma_{r}(E) G^{A}(E)]
\end{equation}
\noindent
where $G^{R/A}(E)$ are the retarded and advanced Green's function of the central region and $\Sigma_{l/r}$ are coupling 
matrices to the left/right leads. Based on the eigenchannel decomposition of the conductance, this total transmission $T(E)$
can be decomposed into nonmixing eigenchannels \cite{Datta,Haug} as follows :

\begin{equation}
T(E) = \sum_{\sigma,n} T_{n}^{\sigma}(E). 
\end{equation}
where $\sigma$ is the spin index. In all our transport calculations, the complex part of the integral leading to
the charge density is computed by using 16 energy points on the complex semicircle, 16 points along the line parallel to
the real axis, and 16 poles. The integral over real energies necessary at finite bias is evaluated at 200 points. 
On the other hand, all the calculations are carried out with periodic boundary conditions in the direction parallel to that of
the transport. In addition, Due to the large size of the unit cell, no k-points are used in the
direction perpendicular to the transport (only the $\Gamma$ point).
    
\section{Results and discussion}
\subsection{Study of the CoPc/Co(111) system}
In order to investigate the interaction between the substrate and CoPc molecule, 
we first perform ab initio calculations to find the equilibrium position of the
CoPc molecule adsorbed on the Co(111) surface. Several initial adsorption 
configurations including hollow, fcc, and bridge sites (see Fig. 2(a)) 
are considered in this sense to find the most stable one. The Co(111) substrate is modeled by
a slab consisting of five layers. For technical considerations, we tested 
two periodic supercells of CoPc/Co(111): $6\times7$  and $7\times8$.
For both supercells, the lattice parameter perpendicular to the surface is 15 \AA,
and each supercell consists of 267 and 337 atoms respectively, including the 57 atoms of CoPc.
We have shown that the two supercells produce adsorption energies 
and electronic structures that are not significantly different from each other.

To obtain the Hamiltonian matrix elements, we used a real space mesh with cutoff of 
500 Ry and a $4\times4$ inplane k-point grid. 
All the geometries are optimized using the conjugate gradient until the 
forces on each atom are smaller than 0.04 eV/\AA. The results are confirmed
by PWscf ab initio simulation package calculations \cite{XiChen}.

Regarding the stability of the CoPc molecule on Co(111), we found that the bridge position 
(see Fig.2(a)) is the most stable and the optimal distance between the molecule and the substrate 
is 2.049 \AA. These results are in good agreement with previous experimental and theoretical
studies \cite{CI,XiChen}. Our calculated relative energy and magnetic moment of the central
atom of CoPc (Co) within the optimized structure of the CoPc molecule adsorbed on Co(111) 
are listed in table 1 for each of the different adsorption sites.

\begin{table}[h]
\begin{center}
\caption{\textit{Calculated relative energy $\Delta E$ (eV) and magnetic moment of Co atom
for different contact geometries. The energy for the most stable position is set to zero.}}
\begin{tabular}{|c|c|c|c|c|c|}
\hline
\hline
                     &  $hcp$     & $fcc$     & $bridge$ \\ 
\hline
$\Delta E(eV)$        &  0.90       &   0.82     & 0 \\
\hline
$\mu(\mu_{B}/atom)$  & 0.52 & 0.46 & 0.34 \\
\hline
\hline
\end{tabular}
\label{tab:tab1}
\end{center}
\end{table}  
 
The difference of relative energy between the hcp and the fcc configurations is small
(80 meV), but it indicates that the fcc configuration is the most favorable. 
In the following, we will present and discuss the results for the most stable configuration.
The magnetic moment (0.34 $\mu_{B}/atom$) of the central atom of CoPc molecule
decreases in comparison to the free molecule value (1.1 $\mu_{B}/atom$).

We also found that a small negative magnetic moment appear for the N and C atoms of CoPc molecule.  
In addition, the magnetic moments of Co atoms on the surface have been modified compared to 
the value (1.85 $\mu_{B}/atom$) of Co atoms of free surface. For example, the magnetic moment 
of each Co atom on the sites A and B (see Fig.2(b)) has a drop of about 0.2 $\mu_{B}$ on average.
In sites C1 and C2, we have an average reduction of about 0.35 $\mu_{B}$ and
0.32 $\mu_{B}$ respectively. 

\begin{figure}[htpb]
\centering
\includegraphics[width=10cm, height=8cm]{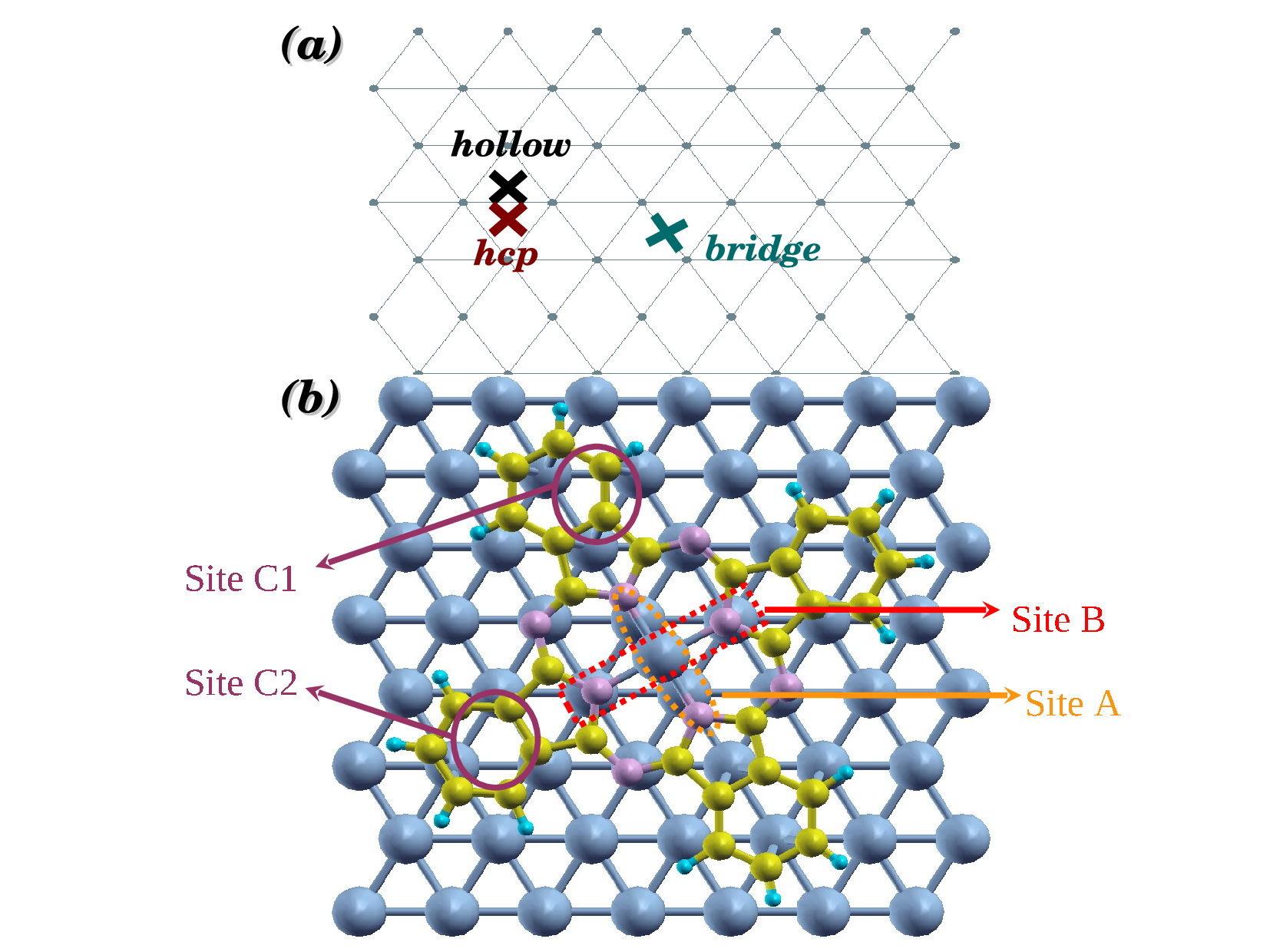}
\label{part-surface-Co}
\caption{\textit{Figure (a) shows the high-symmetry sites of Co(111) surface and the figure (b) shows the 
position of the CoPc molecule centered on the bridge position. Figure (b) shows the atoms of the
Co(111) surface and the molecule CoPc deposited on it.
Each of the sites A and B represent two atoms of Co located just below the nearest four 
nitrogen atoms. In site A, the two Co atoms are in contact with the central atom and two N atoms, 
whereas the two atoms of site B are only in contact with two nitrogen atoms. Site C1(2) represents 
the Co atoms forming a triangle below the benzene ring.}}
\end{figure}

Indeed, these changes of the magnetization can be also observed from the density of states.
Figure 3 represents the PDOS of the Co atom of the molecule, those of the nearest four N atoms,
and the surface Co atoms. In this Figure (3), we can observe that there is a hybridization between    
the Co surface atoms and N atoms at binding energies between -4.7 eV and -3.8 eV.
This figure shows also a hybridization between Co atom of CoPc with the Co(111) surface in low energy range 
(between -5.6 and -6.3 eV, -7.5 and -8.1 eV, -9.5 and -9.7 eV). 
This hybridization provides the chemical bonding of the CoPc on the magnetic surface.
In addition, the PDOS of the Co atom of the CoPc molecule shows a shift of the up (down) band 
to the right (left), which explains the reduction of the magnetic moment compared to this one 
found in the case of the free molecule.

\begin{figure}[htpb]
\centering
\includegraphics[width=9cm, height=11cm, angle=-90]{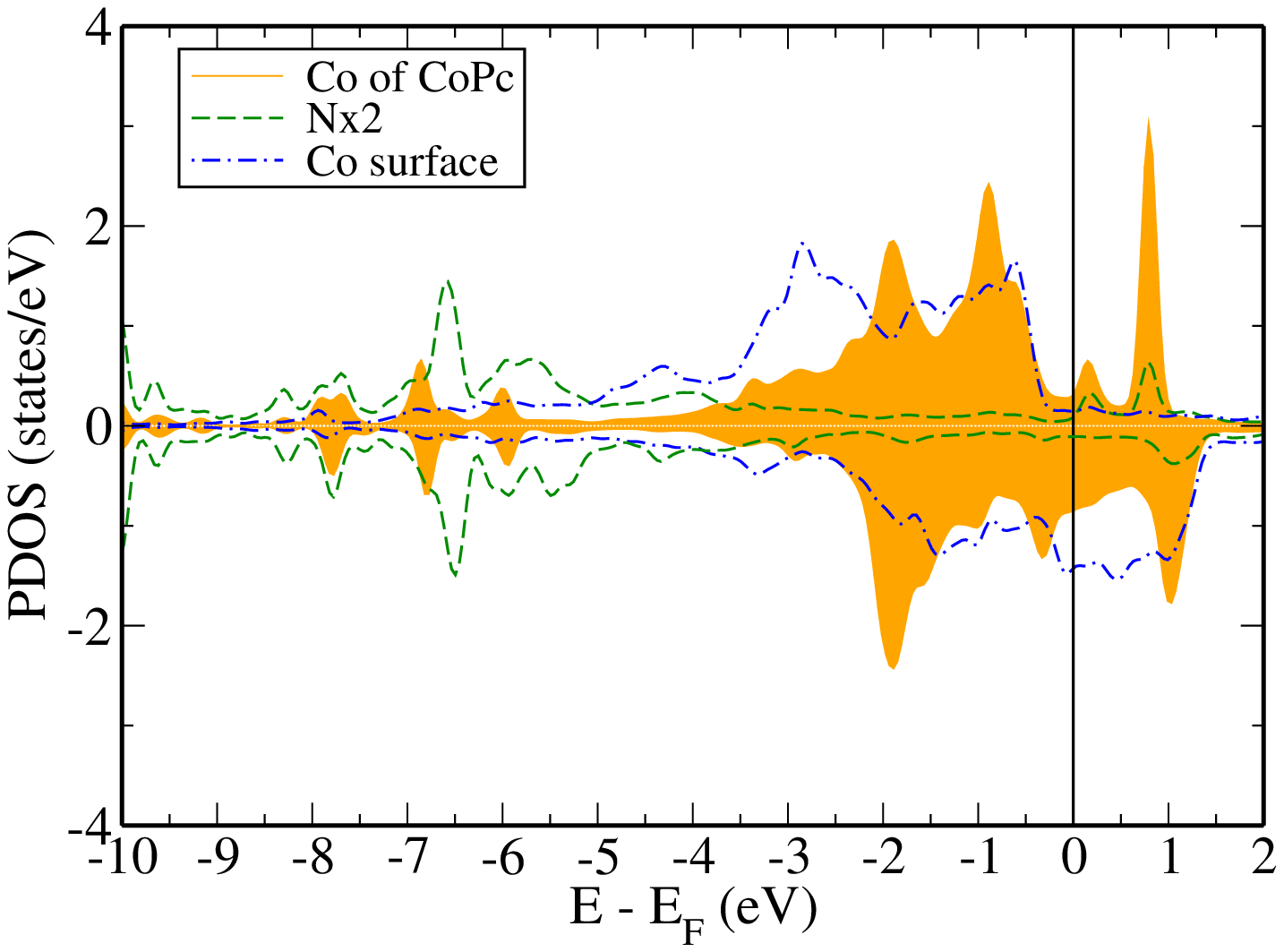}
\caption{Calculated spin-polarized PDOS of various atoms of CoPc/Co(111).}
\label{fig:3}
\end{figure}

Consequently, those modifications, that were induced on the electronic and magnetic properties 
of the surface's Co atoms and the atoms of the CoPc molecule, are the sign of a pronounced 
interaction between CoPc and the magnetic substrate. Here we should point out that our values 
are in good agreement with others obtained in previous studies by means of other methods \cite{CI,XiChen}.

\subsection{Electronic and magnetic properties of the STM-tip in contact with CoPc/Co(111)}
To explore the transport properties as function of the distance between the STM Co-tip and the substrate in
the CoPc molecular junction, we have first investigated the electronic and magnetic properties of the two
configurations (see Fig.1 (a) and (b)). We placed the STM-tip above the Co atom of the CoPc molecule, and considred
the range of the varying distance between the Co-tip and the CoPc between 2.05 and 6.0 \AA. 

\begin{figure}[htpb]
\centering
\includegraphics[width=14cm, height=13cm]{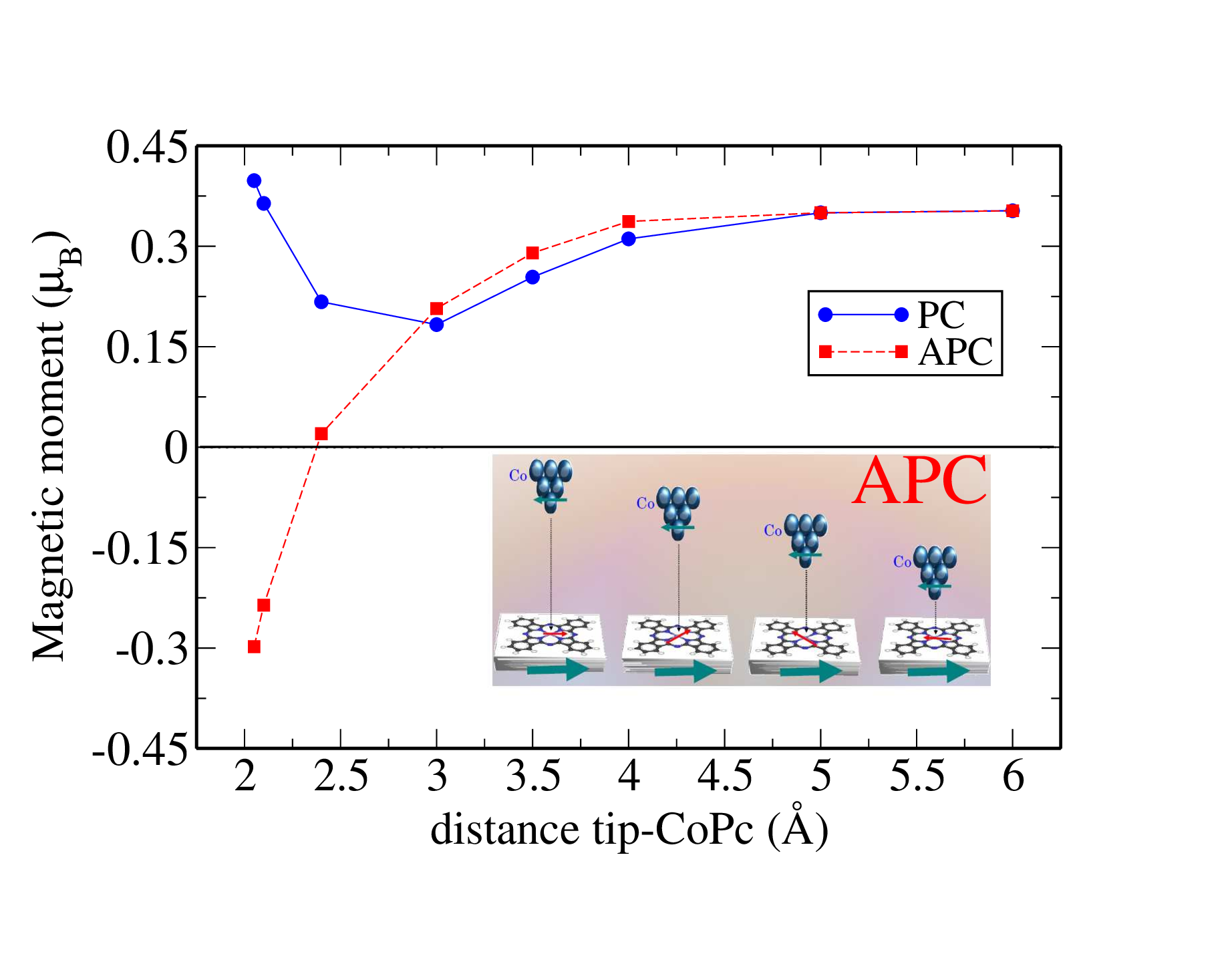}
\caption{Magnetic moment of Co atom's of CoPc molecule as a function of the tip-CoPc distance.}
\label{fig:4}
\end{figure}

This magnetic moment for both configurations is ploted in Fig.4, in a range which varies
from the tunneling to the contact regime. For the parallel configuration (PC) the magnetic 
moment is always positive. We can observe that the value of the magnetic moment at 
5-6 \AA{}, where the interaction of tip-CoPc is weak, is similar to that found in
the case where the molecule is deposited on Co(111). At 2.05 \AA{}, where the interaction
between the tip and CoPc is strong, the value of the magnetic moment is larger than that
obtained at 5-6 \AA{}. As for the antiparallel configuration (APC), the magnetic moments
observed in the range 3-6 \AA{} are almost the same as those in the parallel configuration, 
since we are in the tunneling regime. The appealing result in the APC is the spin flip which
starts at 2.4 \AA{} where the magnetic moment of the central atom is almost zero.   
This modification of the magnetic moment depending on the distance between tip-CoPc
is a well-known effect \cite{R.Z.Huang,Tao} and is attributed to the weakening of the 
hybridization-interaction between the tip and an adatom-surface.
 
To explain the origin of the spin flip, we need to understand the hybridization-interaction 
from the electronic structures of the s and $d$ bands. For this purpose, we analyzed the charge distribution
of the $s$ and $d$ bands and the density of states of the central atom. In order to make a
clear comparison, we choose to present the charge distribution and the density of state at 2.05 \AA{}
(contact regime), 2.4 \AA{} (point of transition) and 5.0 \AA{} (tunneling regime)
for both configurations. The charge distributions of $s$ and $d$ bands are plotted in figure 5, and the total
DOS of the central atom of CoPc is presented in figure 6.

\begin{figure}[htpb]
\centering
\includegraphics[width=12cm, height=14cm, angle=-90]{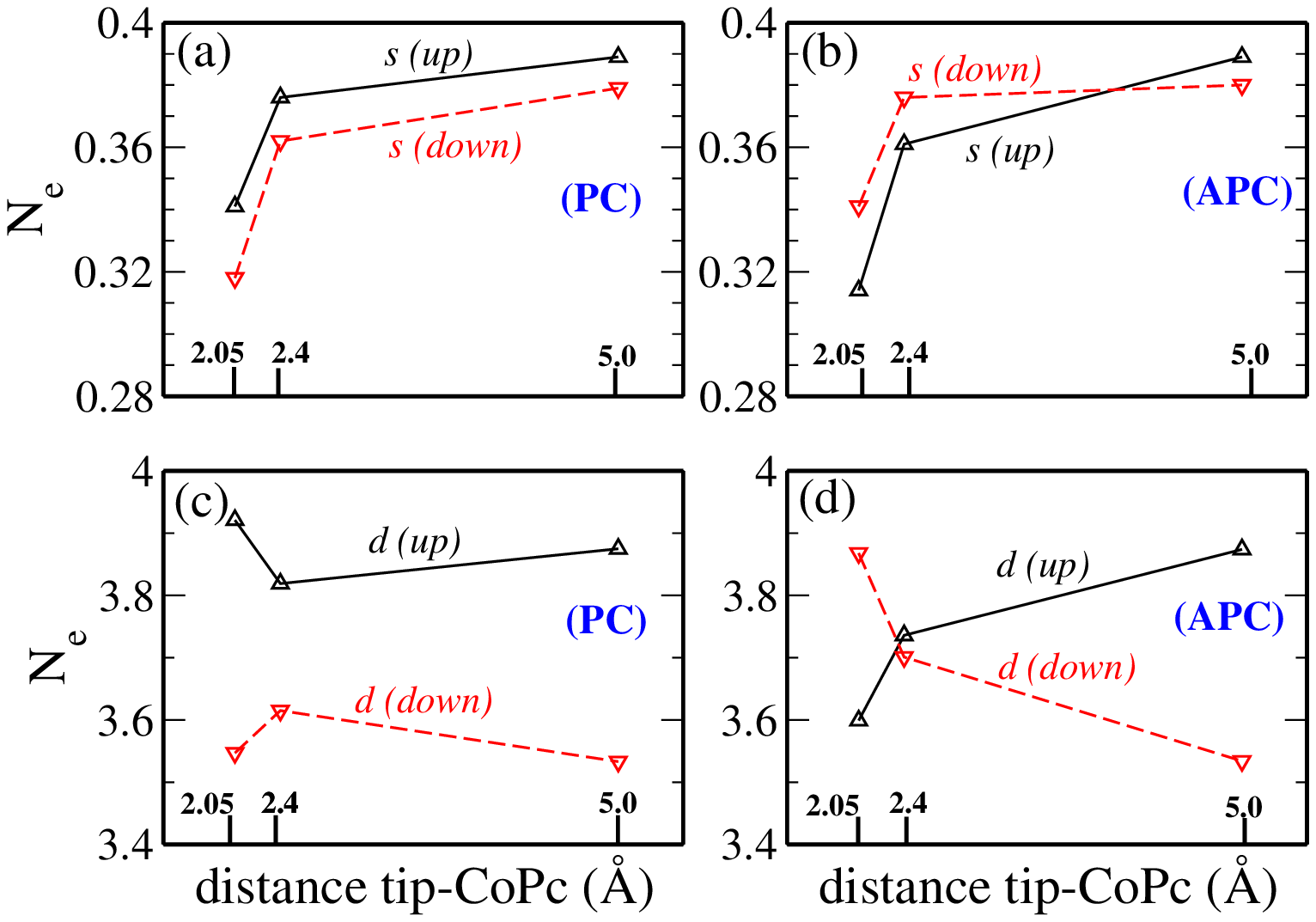}
\caption{Charge distribution as a function of the tip-CoPc distance, for each spin orientation of the Co in CoPc in the parallel
(left) and anti-parallel (right) configuration.}
\label{fig:5}
\end{figure} 

The figure 6 shows that the main contribution is obviously from the $d$ band.
For the $s$ band, the difference between up and down charge is very small in the 
two configurations (PC (Fig.5-a), and APC (Fig.5-b)). However, it is positive in the PC and negative in APC
in contact regime where the spin flips. For the $d$ band, our analysis of the charge distribution 
for the APC (Fig.5-d) reveals that the tip-CoPc interaction depletes the \textit{up-d} states of the Co in CoPc, 
and increases the population of the \textit{down-d} states which becomes the majority in the contact regime ($d=2.05$ \AA{}). 
We can also observe this effect by comparing of the DOS of the Co in CoPc, in
the three chosen tip-CoPc separations, with that calculated in the CoPc/Co(111) system (see Fig.6 (a) and (b)).
The results for both spin directions are plotted where the energies are given relative to
the Fermi energy. In the parallel configuration (Fig.6(a)), both the $up$ and $down$ 
spin states show the same negative shift. It can be easily observed that the shift of
the two states are affected by the tip-substrate distance. However, these shifts increase when 
the tip-CoPc distance decreases, i.e. the interaction becomes stronger. At $d = 5.0$ \AA{},
where the interaction of the tip with the Co atom in CoPc is rather weak, the DOS of the 
central atom is very similar to the case where the CoPc is deposited on Co(111). 
Similary, in the anti-parallel configuration (Fig.6(b)), the $up$ and $down$ states are 
shifted negatively, and these shifts depend on the tip-CoPc distance.
However, it is important to point out that these states are not shifted in the same way. 
The shift of the spin $up$ states is smaller compared with that of the spin $down$ states. 
This explains the observed transition of the magnetic state from ferromagnetic to anti-ferromagnetic. 
Finally, we compare in figures 6(c) and 6(d) the DOS of the central atom determined in both configurations
(PC and APC) in contact ($d=2.05$ \AA{}) and tunneling ($d=5.0$ \AA{}) regime respectively. 
In the contact regime (Fig.6(c)), we can see clearly that the spin states up (down) the majority, 
which implies a positive (negative) magnetic moment in the PC (APC). In the tunneling regime (Fig.6(d)), 
we observe that the difference between the DOS found in the PC and one found in the APC is negligible.
In addition, we note that the spin $up$ states are the majorities, for that we found in this regime a
positive magnetic moment in both configurations. 
 
\begin{figure}[tbp]
\setlength{\unitlength}{1cm} \vspace{0truecm} 
\centerline{\includegraphics[width=0.35\linewidth,clip,angle=-90]{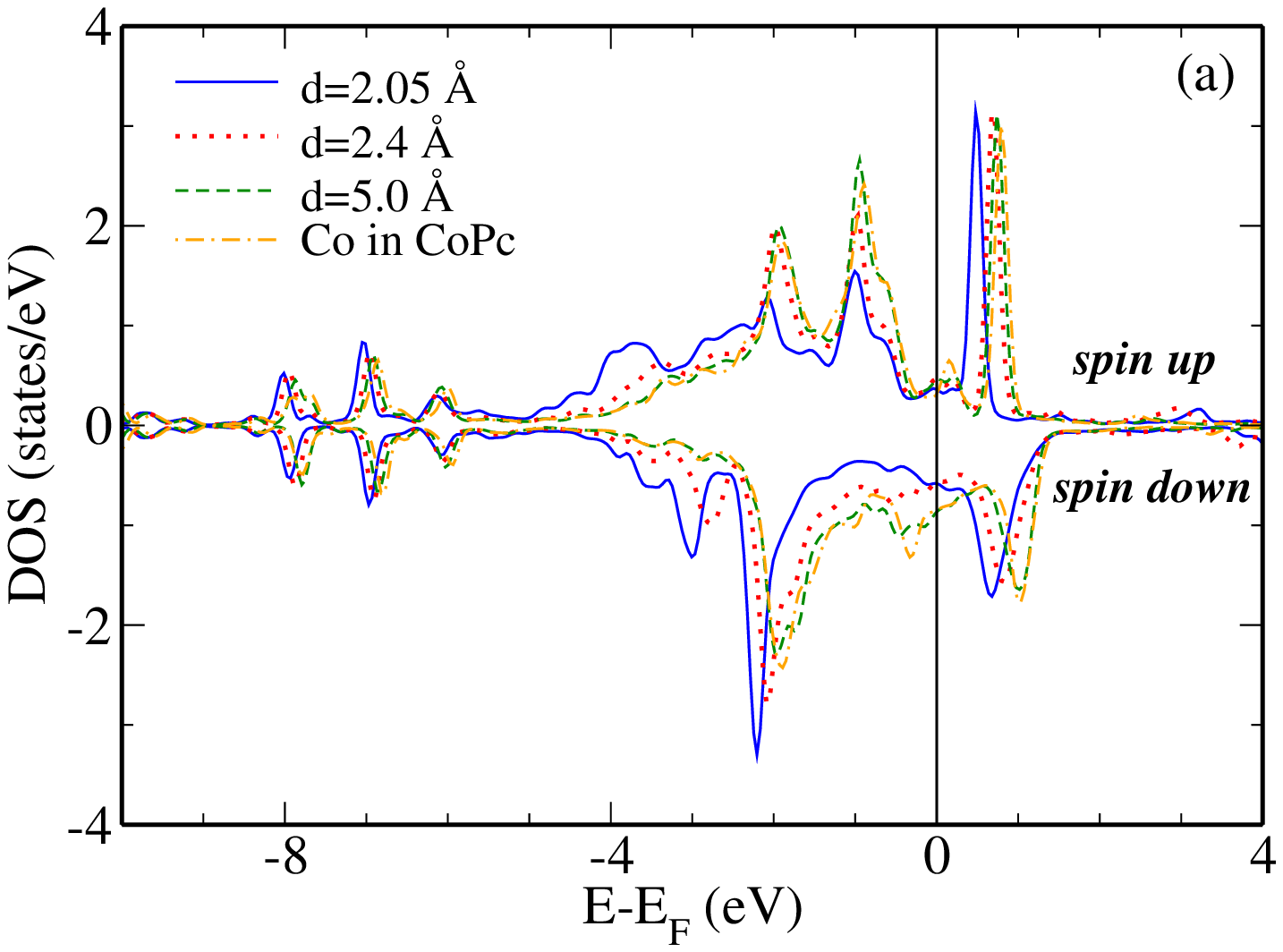}
\includegraphics[width=0.35\linewidth,clip,angle=-90]{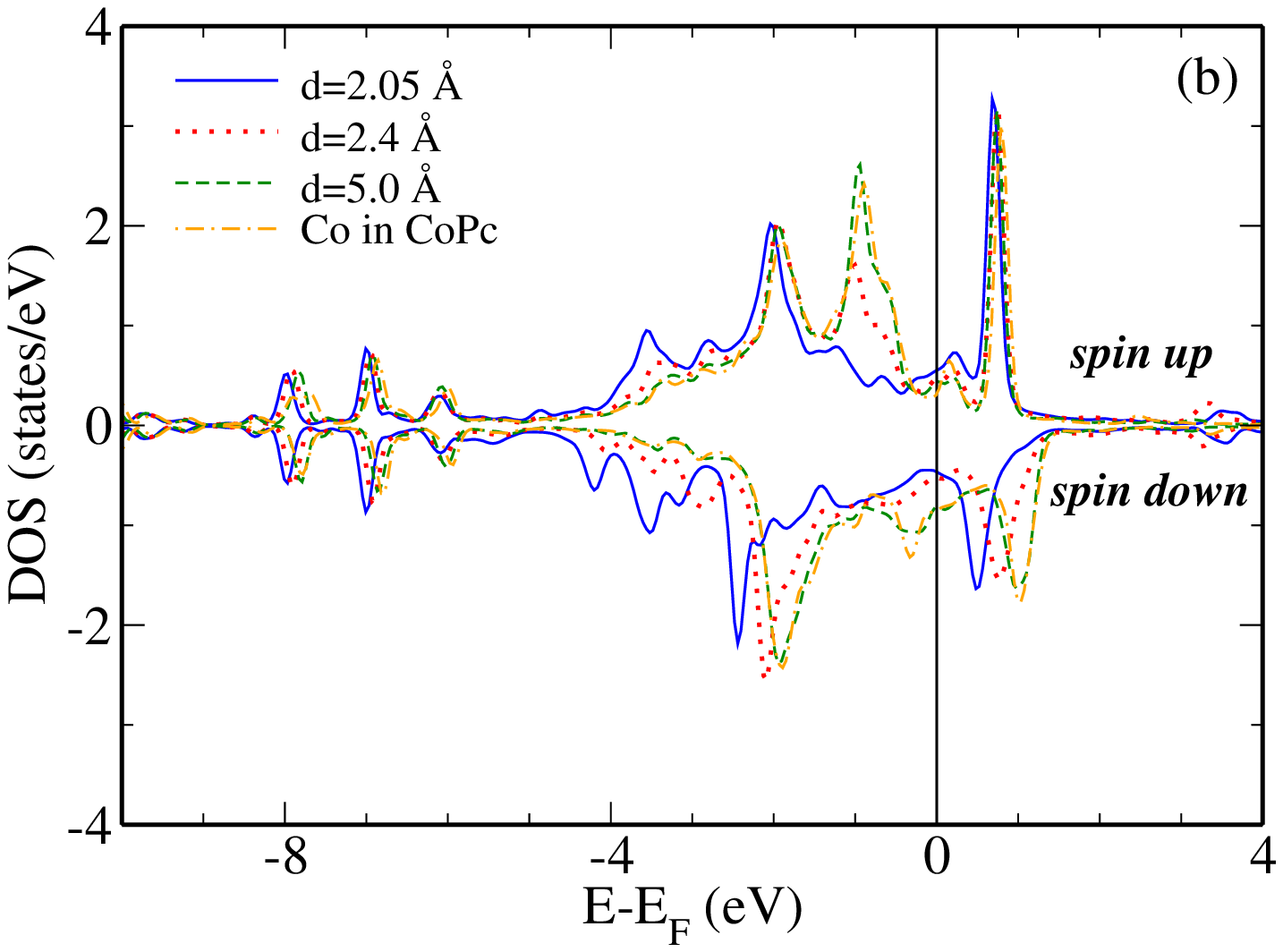}}
{\includegraphics[width=0.35\linewidth,clip,angle=-90]{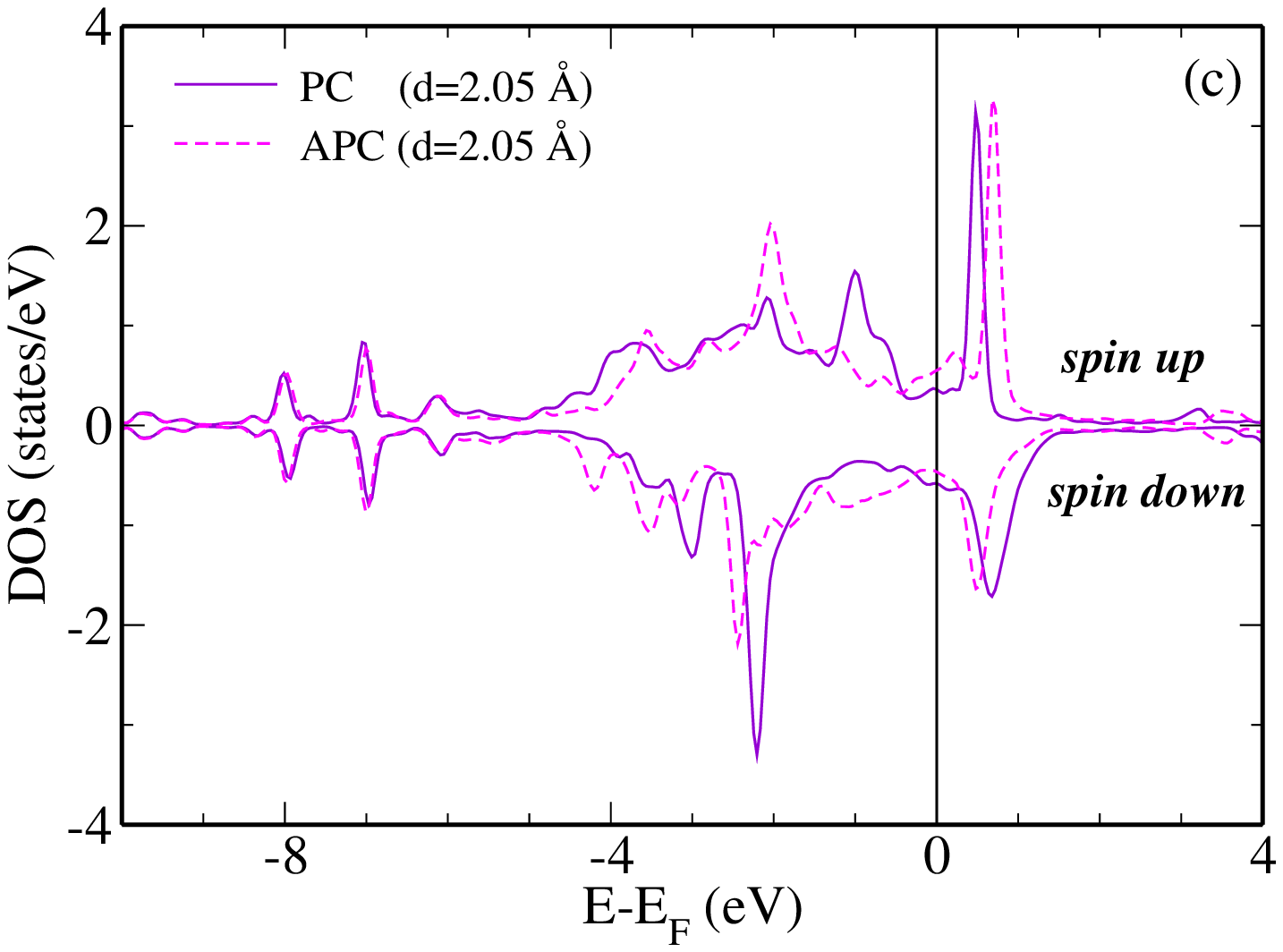}
\includegraphics[width=0.35\linewidth,clip,angle=-90]{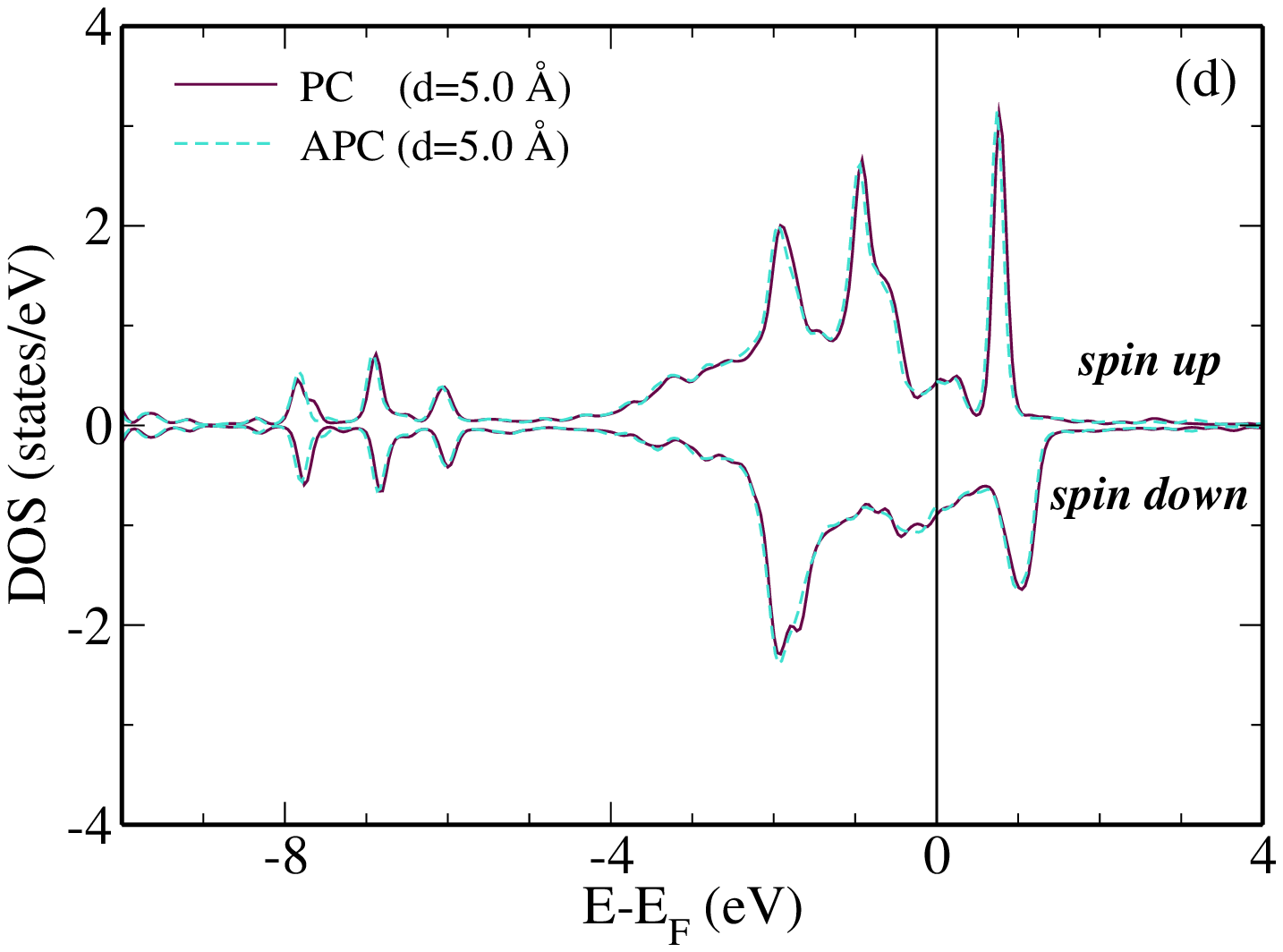}}
\caption{Density of states of the Co in CoPc in the parallel (a) and anti-parallel (b) configurations. (c) and (d) show 
the DOS for both configurations (PC and APC) at 2.05 \AA{} and 5 \AA{} respectively.} \label{fig:6}
\end{figure}

\subsection{Transport properties}
The main conclusion of the previous section is that the interaction between the Co-tip 
and CoPc molecule plays an important role in the modification of the magnetic moments 
of the central atom in CoPc. At this stage, one wonders what is the possible influence of this
interaction on the transport properties. For this aim, we have first performed the transport 
calculations at zero bias. The transmission coefficients calculated in the tunneling  
(d=5.0 \AA{}), transition (d=2.4 \AA{}) and contact (d=2.05 \AA{}) ranges for the parallel
and anti-parallel are plotted in figure 7. In the tunneling regime ($d=5$ \AA{}) where the interaction
is weak, we observe that the transmission coefficients are very small for PC and APC. 
At $d=2.05$ \AA{} and $d=2.4$ \AA{} in the PC (Fig.7(a)), the transmission at the Fermi level of 
the spin $up$ is almost the same in all ranges, and the transmission of the spin $down$ increases
when the tip-CoPc separation decreases. In contrary, in the anti-parallel configuration (Fig.7(b)), 
the transmission of the spin $up$ increases. One can then deduce that increasing the separation 
leads to a reduction of the transmission in the two configurations. Therefore, this indicates that 
the electron transfer rate between the tip and CoPc in the contact regime is larger than that in the
other cases. In order to describe the effect of the interaction between the tip and CoPc on electron 
transport at zero bias, we show in figure 8 the variation of the conductance upon approaching
a Co tip to the central Co atom in CoPc/Co(111), in the PC (Fig.8-(a)) and APC (Fig.8-(b)). 
At zero bias the conductance, G, is simply $G=e^{2}/hT(E_{F})$, where $e$ is the electron
charge, $h$ the Planck's constant, and $T(E_{F})$ is the transmission coefficient calculated at $E_{F}$.

\begin{figure}[tbp]
\setlength{\unitlength}{1cm} \vspace{0truecm} 
\centerline{\includegraphics[width=0.35\linewidth,clip,angle=-90]{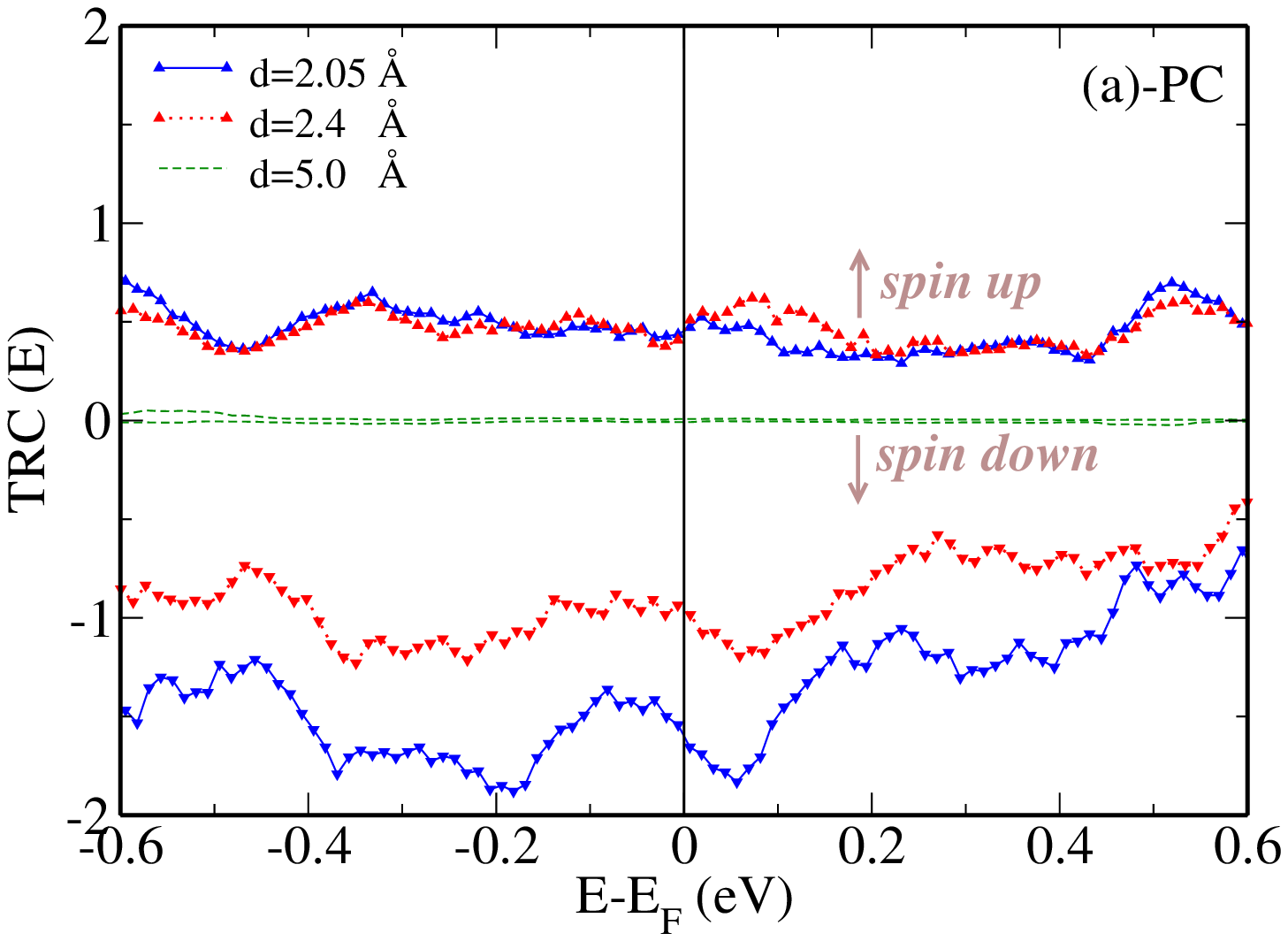}
\includegraphics[width=0.35\linewidth,clip,angle=-90]{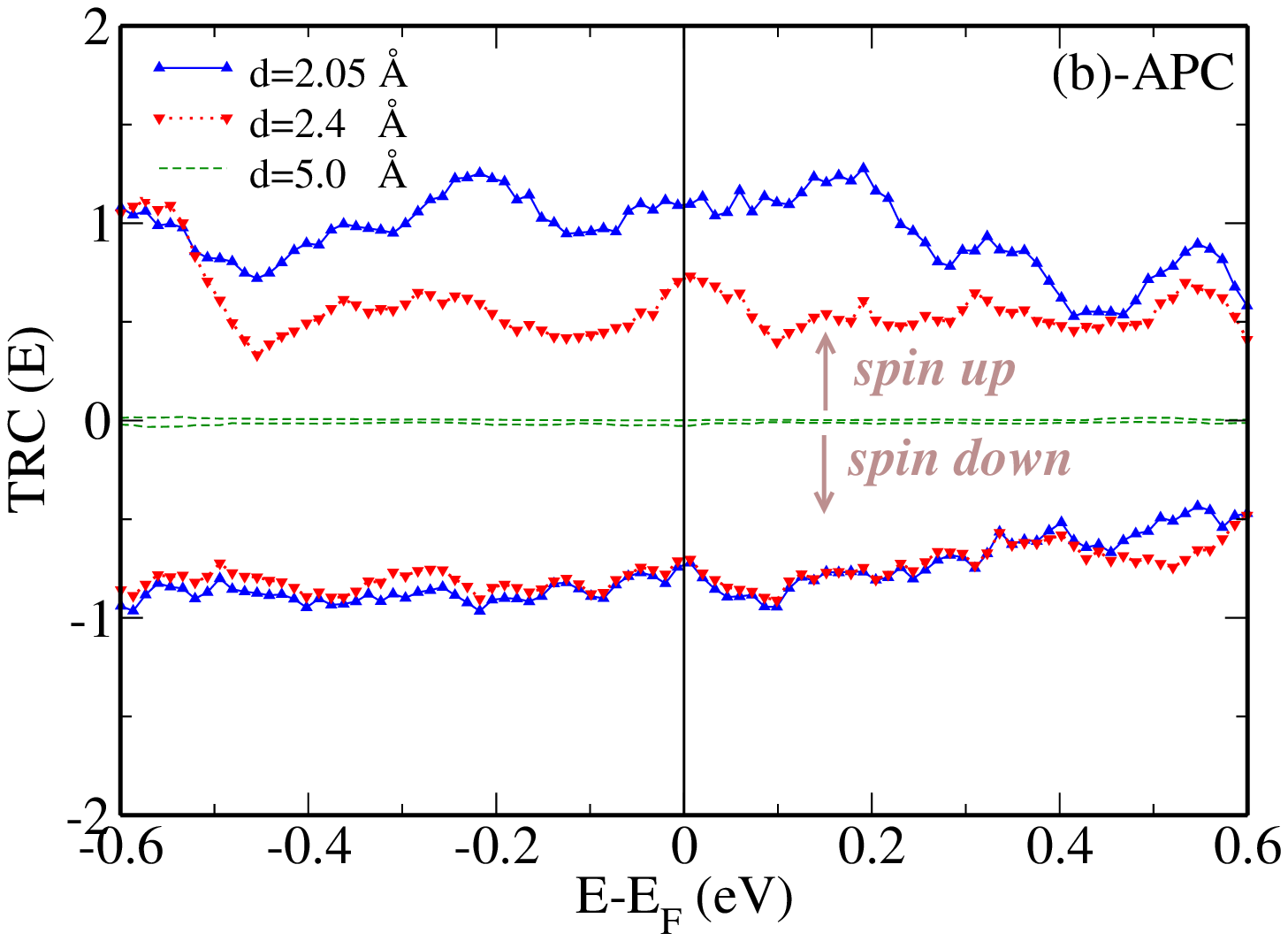}}
\caption{Zero bias transmission coefficient for the tip and CoPc at different tip-CoPc separations ($d$),
(a) for parallel configuration and (b) for anti-parallel configuration. Positive values are for spin up,
negative values for spin down.} \label{fig:7}
\end{figure}

\begin{figure}[htpb]
\centering
\includegraphics[width=8cm, height=11cm, angle=-90]{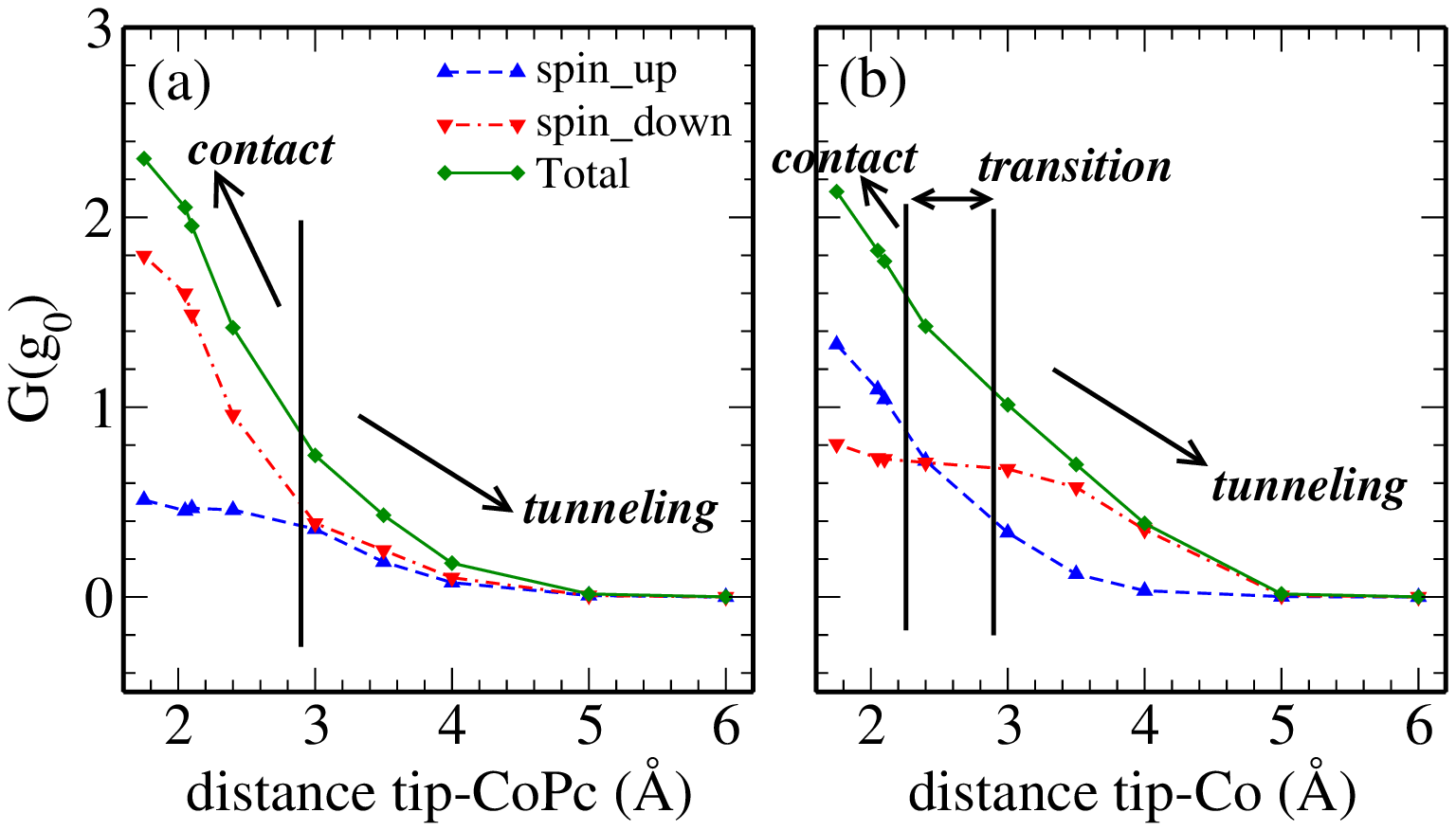}
\label{conduct}
\caption{Conductance in units of $g_{0}$, i.e. the transmission values at Fermi level, as a function of 
the tip-CoPc distance in the parallel (a) and anti-parallel (b) configurations.}
\label{fig:8}
\end{figure}

In the parallel configuration (PC), the conductance  of the spin $down$ states (Fig.8(a))
is larger than that of the spin $up$ states in the contact regime whereas they are almost
similar in the tunneling regime. In the anti-parallel configuration (APC), we observe
that the conductance of spin $up$ states (Fig.8(b)) which dominate till the 2.4 \AA{} 
(contact regime). After this distance (2.4 \AA{}) we can see that the conductance of spin 
$down$ states dominate because the CoPc molecule rotates its spin.
Such explanation can be drawn from the analysis of the orbital projected density of states 
PDOS of the tip atom and the central atom in CoPc. As the main contribution is obviously 
from the $d$ band, we show in the figures 9 and 10 the PDOS (projected density of states)
of $3d$ orbitals of the Co-tip and the Co atom of CoPc respectively.
For the Co-tip atom, figure 9 shows that the spin $down$ ($up$) states dominate around the Fermi 
level in the PC (APC). As for the central atom of CoPc, the figure 10 shows that the majority 
and minority states depend by the tip-CoPc interaction. Besides, this figure shows that the main 
contribution to the PDOS of the majority and minority states is from $d_{xz}$, $d_{yz}$ and $d_{z^{2}}$,
which therefore participate the most to the transmission.
 
\begin{figure}[tbp]
\setlength{\unitlength}{1cm} \vspace{0truecm} 
\centerline{\includegraphics[width=0.4\linewidth,clip,angle=-90]{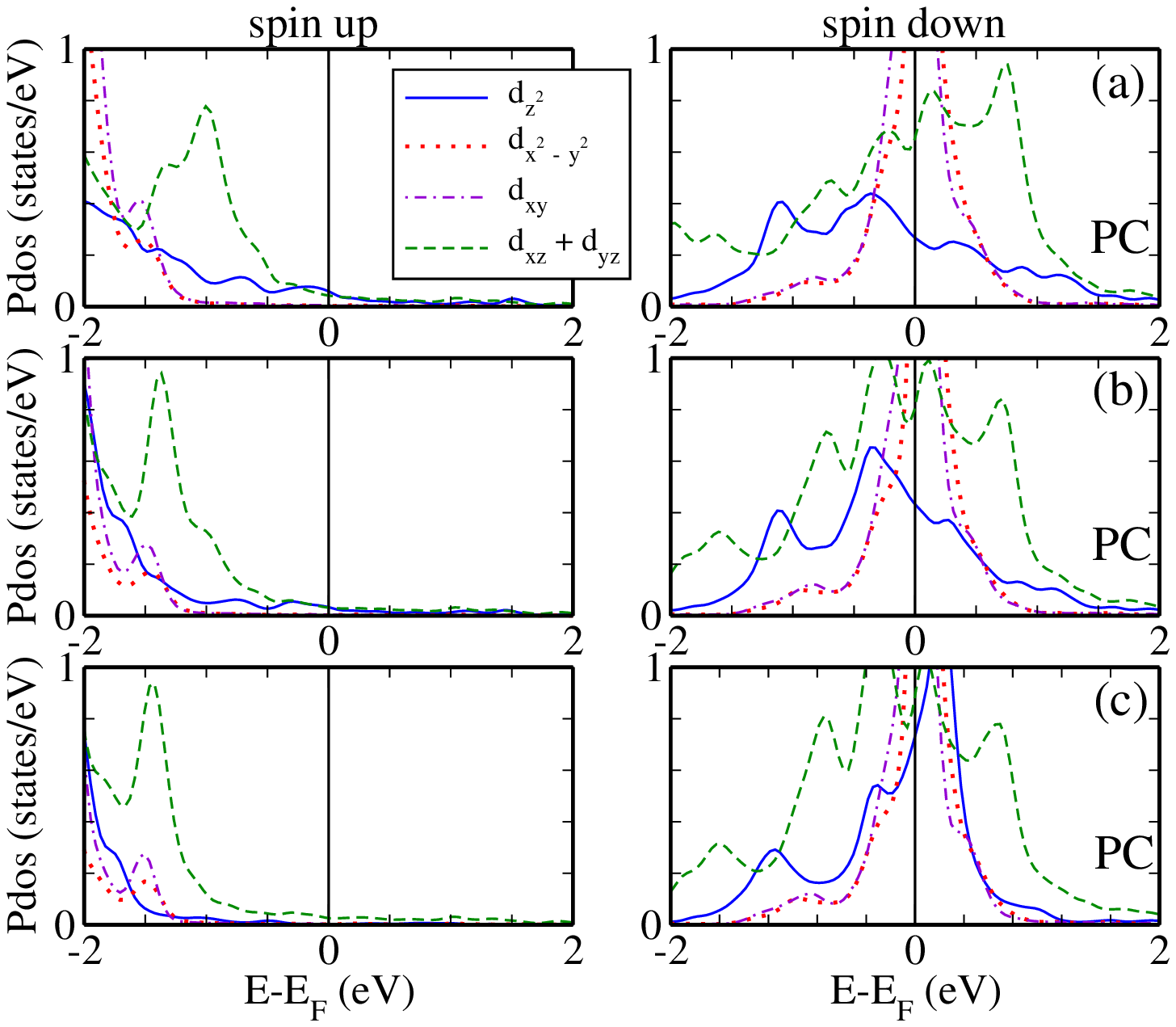}
\includegraphics[width=0.4\linewidth,clip,angle=-90]{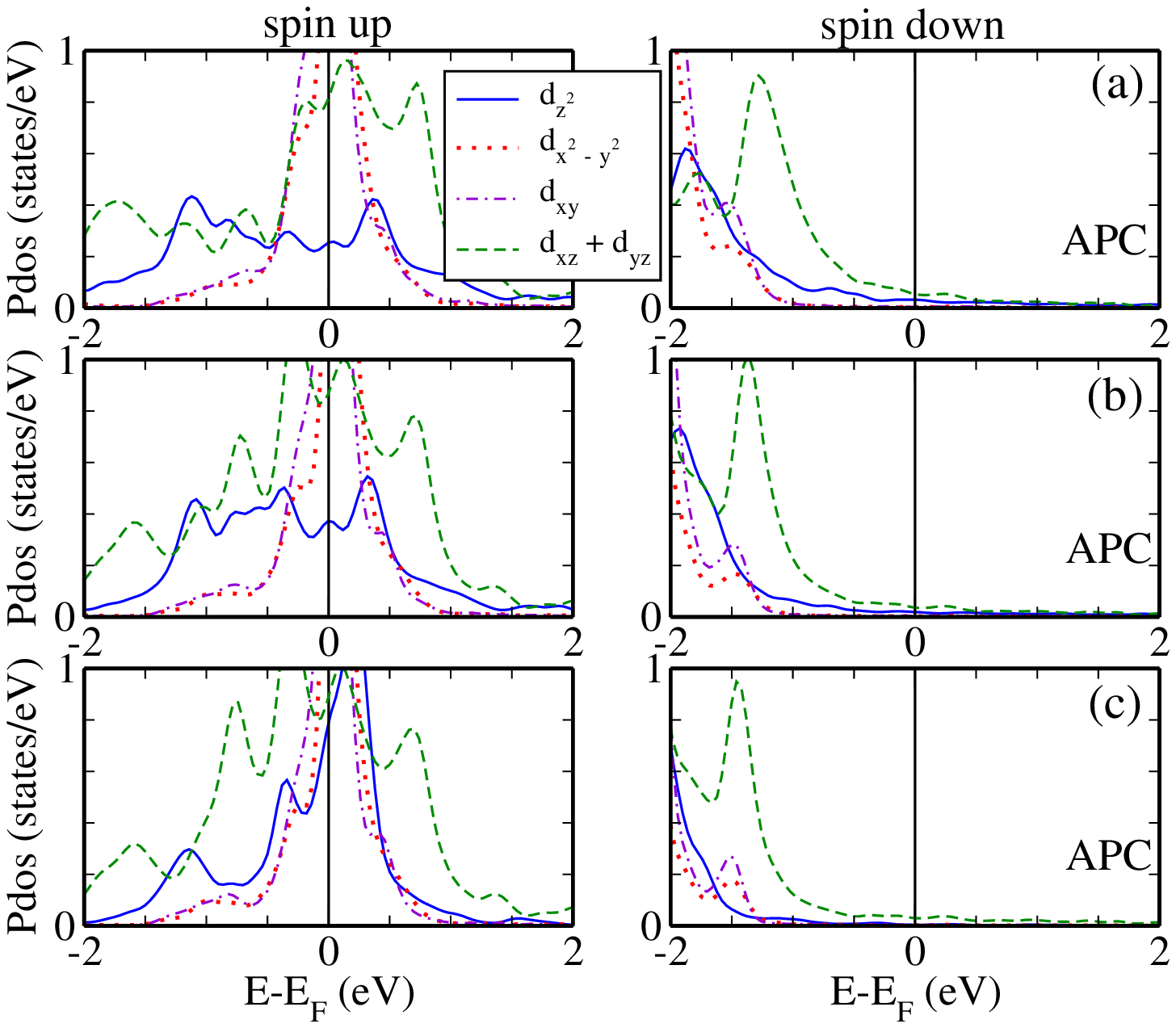}}
\caption{Projected density of states of the tip atom in the parallel (left) and anti-parallel (right) configurations.
(a), (b) and (c) represent the PDOS at 2.05, 2.4 and 5.0 \AA{} respectively.} \label{fig:9}
\end{figure}

\begin{figure}[tbp]
\setlength{\unitlength}{1cm} \vspace{0truecm} 
\centerline{\includegraphics[width=0.4\linewidth,clip,angle=-90]{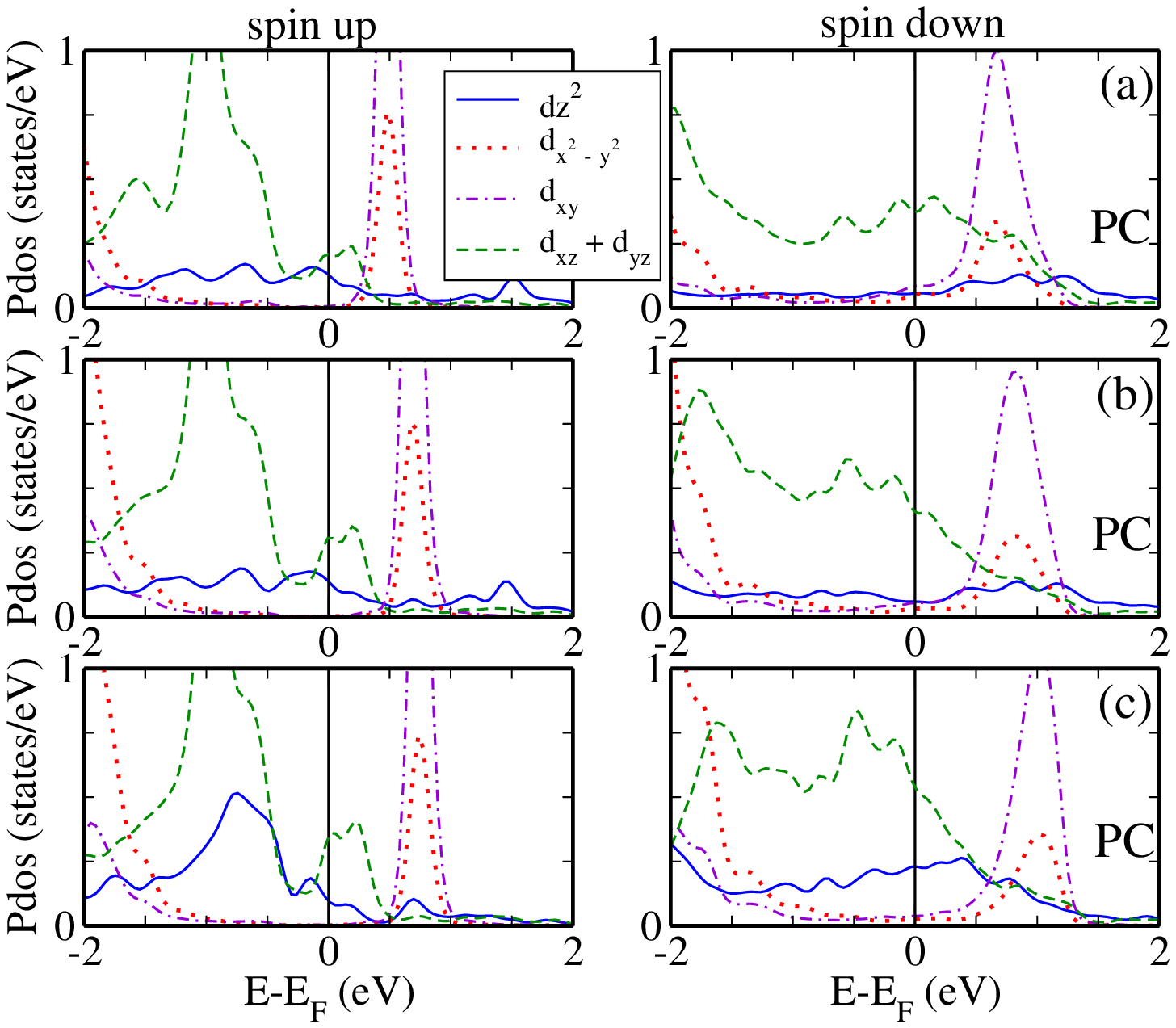}
\includegraphics[width=0.4\linewidth,clip,angle=-90]{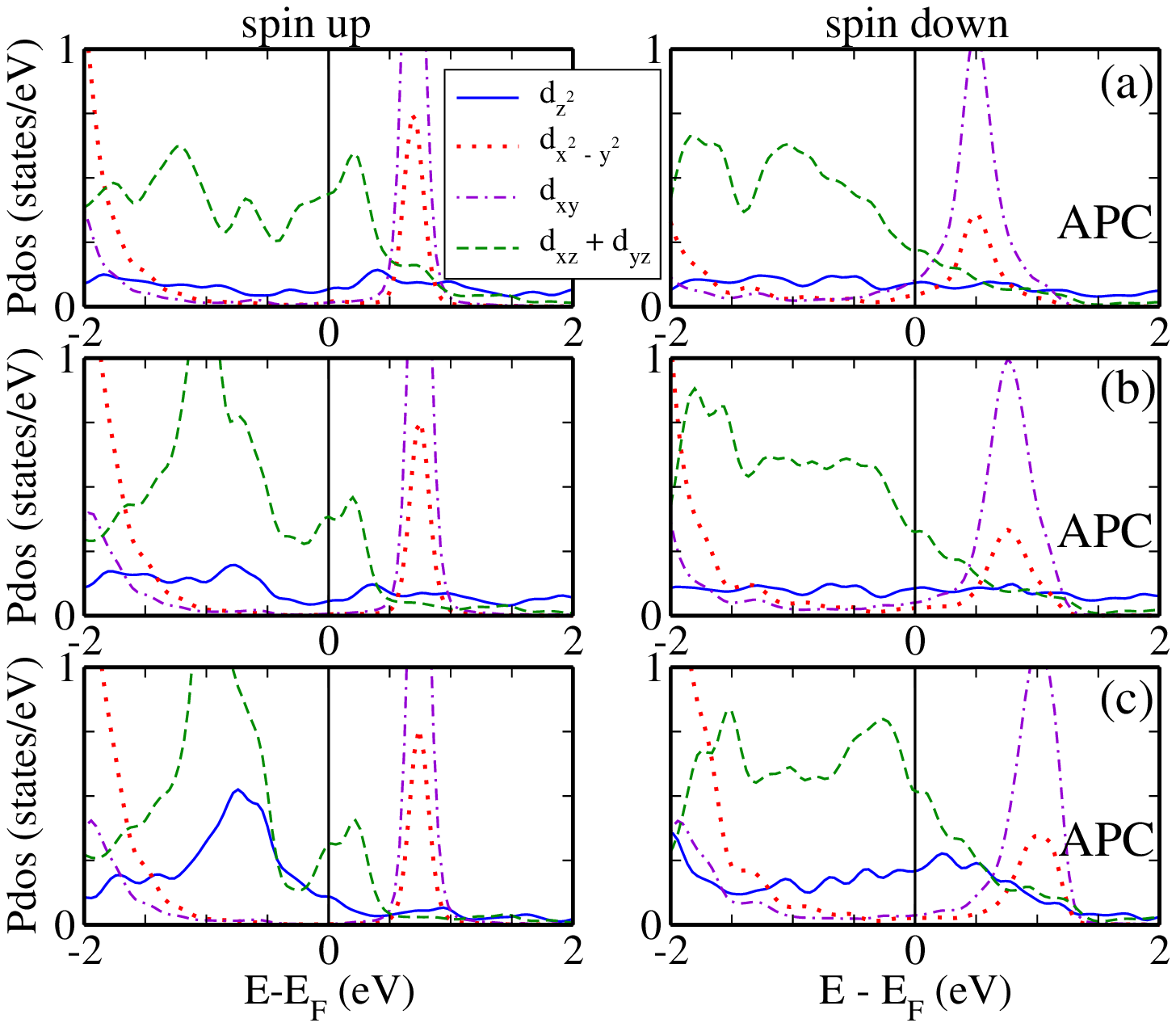}}
\caption{Projected density of states of Co in CoPc in the parallel (left) and anti-parallel (right) configurations.
(a), (b) and (c) represent the PDOS at 2.05, 2.4 and 5.0 \AA{} respectively.} \label{fig:10}
\end{figure}

In the PC (Fig.10, left hand), our calculation show that the contribution of the 
spin $down$ states at Fermi level dominate over the all interval for the Co atom of CoPc.
For this we have that the conductance of the spin-$down$ states is the most important.  
But, in the APC (Fig.10, right hand), we observe that the contribution of the spin $up$ 
states dominate in the contact regime whereas they are minorities in the tunneling regime.
This implies therefore why the total conductance is dominated by the spin-$up$ ($down$)
states in the contact (tunneling) regime. We note that this transition start from the 
distance 2.4 \AA{} where the total contribution of the spin $up$ and $down$ states is nearly the same.
In fact, this transition is due of the strong-weak tip-CoPc interaction. We can also remark,
in the tunneling regime ($d=5.0$ \AA{}), that the contribution of the spin $up$ and $down$ 
states of the central atom (of CoPc) are same in both configurations. This explain why in 
the tunneling regime, the difference between the spin $down$ and spin $up$ conductances is very small.
For the total conductance, obtained by adding the two spin-conductances, we note that the 
values in the contact regime in the parallel configuration are larger than that found
in the antiparallel configuration. These results can be explained by the tip-CoPc interaction
which is more important in the parallel configuration. At 2.4 \AA{} (transition point)
the parallel and antiparallel configurations have almost identical total conductance.
In the other hand, the total conductance in the tunneling regime is always lower than that 
of the antiparallel configuration. Finally, we note that the total conductance, in both
configurations, decreases where the tip-CoPc increase.
   
We can now calculate the $MR$ (\textit{Magneto-Resistance}) as a function of the tip-CoPc separation ($d$) at zero bias, as
$MR (d) = (G^{APC}(d) - G^{PC}(d))/ G^{PC}(d)$. Our calculated values for the $MR$ are shown in 
figure 11. In fact, the changes in conductance lead to very different values for the $MR$. 
In contact (tunneling) regime, we have negative (positive) $MR$, where the 
total conductance $G^{PC} (G^{APC})$ is large of $G^{APC} (G^{PC})$. At transition point the
$MR$ is approximately zero, this is verified by the very samll difference between the total 
 $G^{PC}$ and $G^{APC}$ conductance (see Fig.8). In the tunneling regime, at 3.0 \AA{} (and at 6.0 \AA{})
our calculated value 34.82$\%$ (43.63$\%$) is in good agreement with previous experimental results \cite{CI}.  

\begin{figure}[htpb]
\centering
\includegraphics[width=8cm, height=10cm, angle=-90]{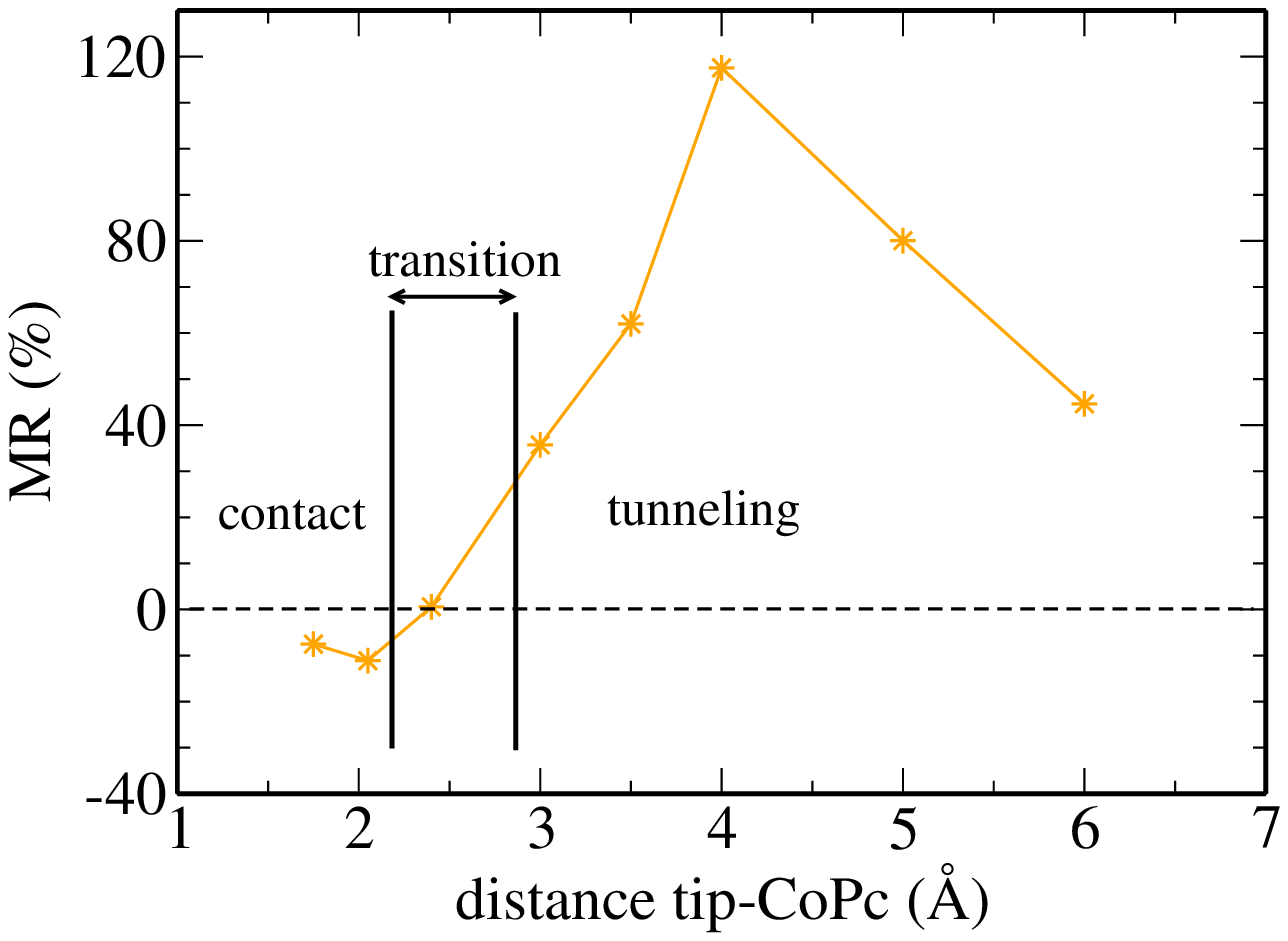}
\label{conduct}
\caption{$MR$ as a function of the tip-CoPc separation at zero bias.}
\label{fig:11}
\end{figure}

\subsection{Bias-dependent current} 
We will analyze now the I-V characteristic. The total tunneling current calculated 
for both configurations in which the spin orientation of the magnetic tip is aligned either 
parallel (PC) or anti-parallel (APC) to that of the magnetic substrate is shown in figure 12
for the contact (fig.12-(a)) and tunneling (fig.12-(b)) regime respectively. The contributions of the total
current of spin $up$ and spin $down$ electrons in both parallel and anti-parallel  
configurations are also shown.
In contact regime the spin $up$ and $down$ currents in the two configurations (Fig.12(a)-PC 
and Fig.12(a)-APC) show an ohmic linear dependence over the voltage, and consequently
for the total current shows the same feature. Unlike the contact regime, in the tunneling 
regime and in both configurations (Fig.12(b)-PC and Fig.12(b)-APC), the spin polarized currents
and the total current present non-ohmic behavior. In addition,
in the tunneling regime, the most remarkable result is that the spin $up$ and spin $down$ 
currents are almost the same in the parallel configuration. However for anti-parallel configuration, 
the spin $up$ current decreases in contrast to an increase of spin $down$ current. 
Besides, the total current obtained in the anti-parallel configuration is larger than that
in the parallel configuration. As for the contact regime, we observe almost similar spin $up$ and spin $down$ currents
in the anti-parallel configuration (Fig.12(a)-APC), in contrast to the behavior in the tunneling
regime (Fig.12(b)-APC). Precisely, in the anti-parallel configuration in the contact regime, we can
distinguish that the spin $up$ current is larger than the spin $down$ current (Fig.12(a)-APC), which
is different than that observed in the tunneling regime (Fig.12(b)-APC). In addition, the total
current for the parallel (Fig.12(a)-PC) and anti-parallel (Fig.12(a)-APC) configurations are almost identical
in the contact regime. 

\begin{figure}[htpb]
\centering
\includegraphics[width=10cm, height=14cm, angle=-90]{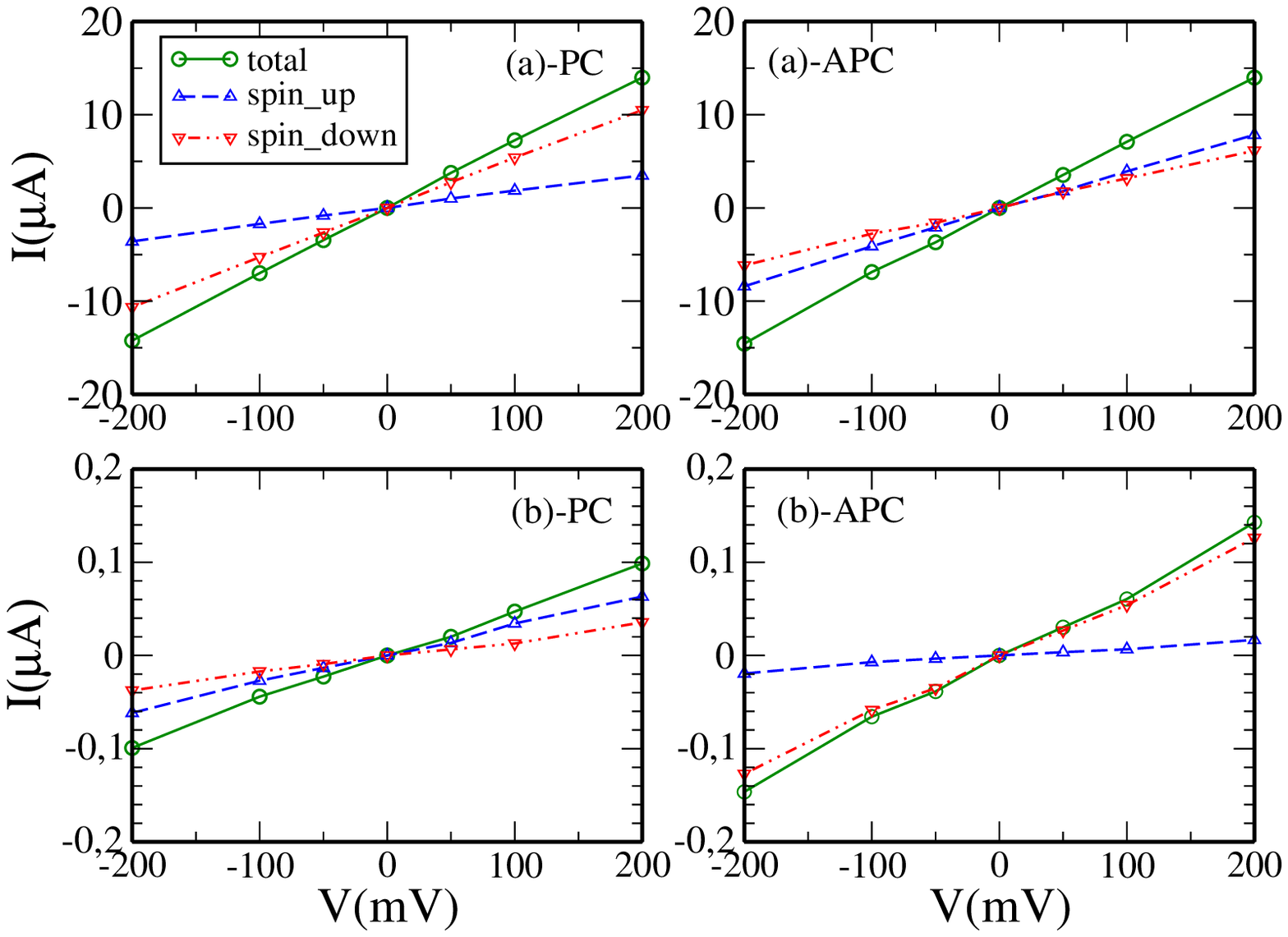}
\label{conduct}
\caption{Spin-polarized current $I$ as function of voltage $V$ for both configurations (PC and APC) 
in contact (a) and tunneling (b) regimes. The tip-CoPc separation is 2.05 \AA{} (5.0 \AA{}) in contact (tunneling) regime.}
\label{fig:12}
\end{figure}

In the case where a positive (negative) bias voltage is applied to the left electrode,
the electrochemical potential of the right electrode is shifted up (down) \cite{Datta1,Datta2},
and hence the density of state at the right molecule-substrate will be shifted up (down). 
Since the transmission coefficient is linked to the density of states at the right 
molecule-electrode interface, the evolution of the transmission spectra will be mainly 
determined by the shift of the electrochemical potential in the right electrode.
In order to analyze our results of \textit{I-V} curves, we discuss the transmission spectra calculated 
at zero bias around fermi level, precisely between -200 $meV$ and +200 $meV$.   

Figure 13 shows the transmission spectra of the CoPc molecule with the Co surface in 
the contact and tunneling regimes. The two \textit{spin-up} and \textit{spin-down} transmissions are 
shown. In the parallel configuration, we can observe that the majority and minority
spin transmissions are almost identical between -0.2 and 0.2 $eV$
in the tunneling regime (Fig.13(b)). However, this difference in the contact regime is more pronounced where   
the \textit{spin-up} transmission is larger than the \textit{spin-down} transmission all over the interval 
[-0.2,+0.2] (eV). These two remarks explain the small and large difference which was observed between the 
\textit{spin-up} and \textit{spin-down} currents in the tunneling (Fig.12(b)-PC) and contact regime 
respectively (Fig.12(a)-PC). On the other hand, the transmission coefficient for the anti-parallel 
configuration, shows that the \textit{spin-down} transmission becomes larger than that of \textit{spin-up}
in the tunneling regime (Fig.13(b)). We can also observe that the \textit{spin-up} and \textit{spin-down}
transmissions become almost similar in the contact regime (Fig.13(a)). 
The two observations explain why the total current in the tunneling regime is dominated by the 
\textit{spin-down} current (Fig.12(b)-APC), and that the two \textit{spin-up} and \textit{spin-down}
spin currents are almost similar in the contact regime (Fig.12(a)-APC).

\begin{figure}[tbp]
\centering
\setlength{\unitlength}{1cm} \vspace{0truecm} 
\centerline{\includegraphics[width=0.4\linewidth,clip,angle=-90]{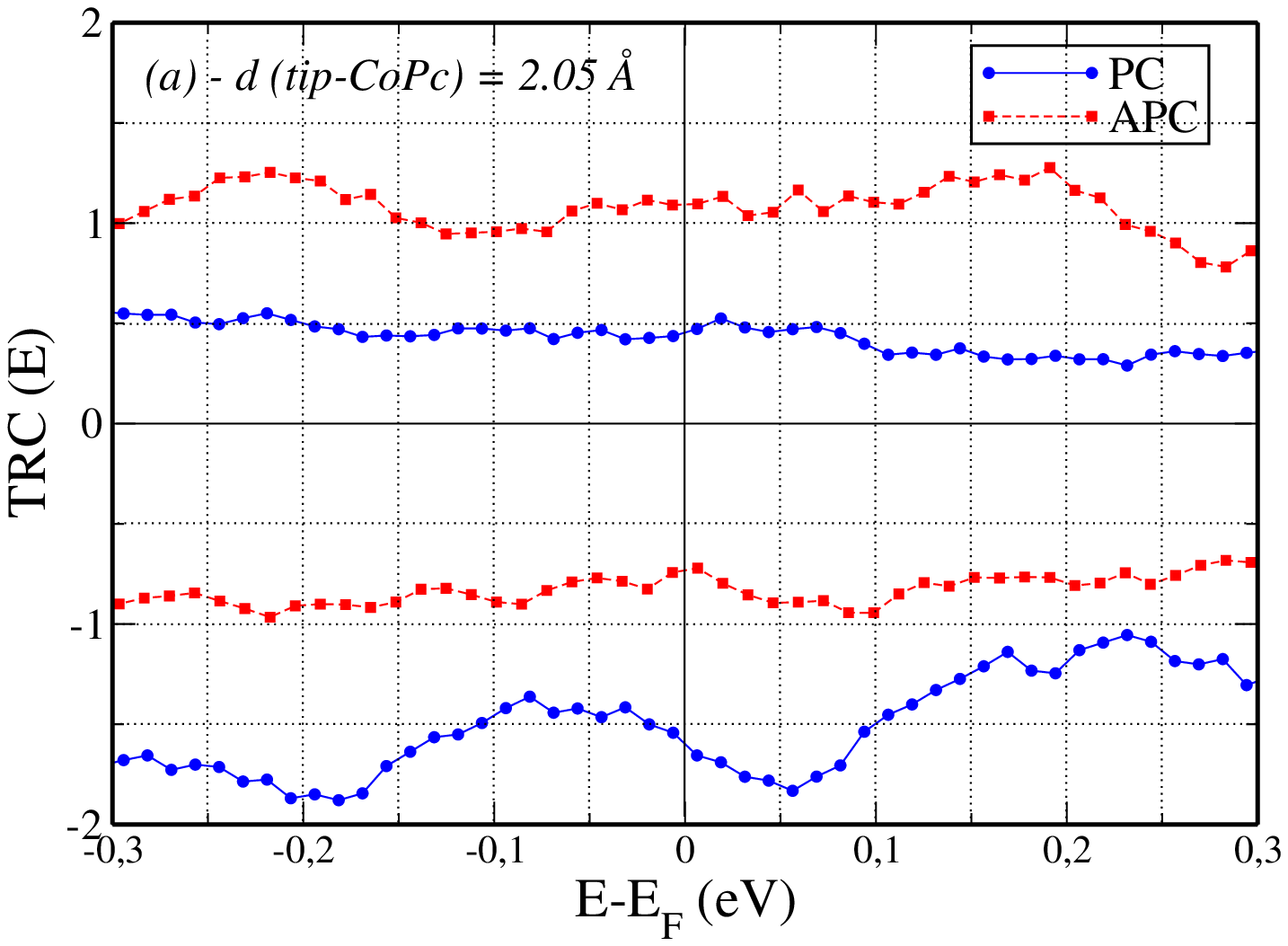}
\includegraphics[width=0.4\linewidth,clip,angle=-90]{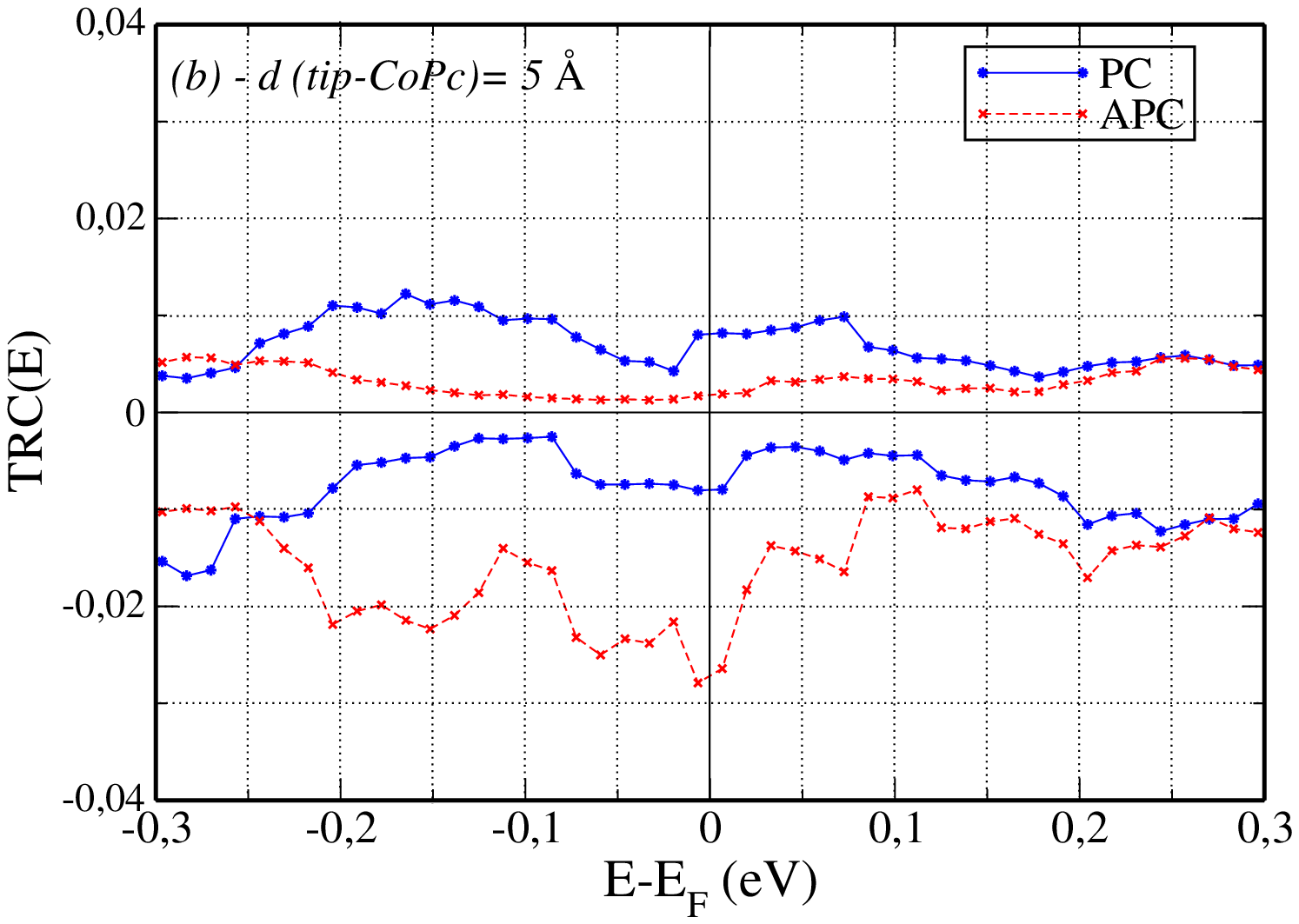}}
\caption{\textit{Zero bias transmission coefficient for parallel and antiparallel configurations 
in contact (at distance 2.05 \AA{}) and tunneling (at distance 5 \AA{}) regimes. Positive values are for spin-up, negative
values are for spin-down.}} \label{fig:13}
\end{figure}

\section*{Acknowledgments}
One of us (T.K.) would like to thank Campus France and the French Embassy at Lebanon for the scholarship during his scientific stay in France(program SAFAR/SSHN) as  well as the Laboratory of the Institut des Mol\'ecules et Mat\'eriaux du Mans(IMMM) of the Le Mans  Universit\'e (CNRS-UMR 6283 ULP) for its kind hospitality during the time the present work was completed.\\

\end{document}